\documentclass[sigconf, screen]{acmart}

\newcommand{\eg}[0]{\textit{e.g., }}
\newcommand{\ie}[0]{\textit{i.e., }}

\newcommand{\revision}[1]{{\color{black} #1}}
\newcommand{\revisionNew}[1]{{\color{black} #1}}

\newcommand{\ours}{\texttt{\tool{}}}

\usepackage{comment}
\usepackage{stfloats}
\usepackage{microtype}
\AtBeginDocument{%
  \providecommand\BibTeX{{%
    \normalfont B\kern-0.5em{\scshape i\kern-0.25em b}\kern-0.8em\TeX}}}

\usepackage[autostyle,german=guillemets]{csquotes}
\makeatletter

\usepackage{hyperref,quoting}
\usepackage{url}
\quotingsetup{vskip=0pt,font={itshape,raggedright},rightmargin=0pt}
\usepackage{tikz}
\usepackage{cuted}
\usepackage{capt-of}
\usepackage{subcaption}
\usepackage{fancyvrb}
\usepackage{gensymb}
\usepackage{float}
\usepackage{stfloats}
\usepackage{tikz}
\usepackage{xcolor}
\definecolor{purple}{RGB}{153, 102, 255}
\newcommand{\purplesquare}[1]{%
    \begin{tikzpicture}[baseline=(char.base)]
        \node[draw=purple, fill=purple, rectangle, rounded corners=2pt, minimum height=10pt, minimum width=10pt, inner sep=0pt, text=white] (char) {\textbf{#1}};
    \end{tikzpicture}%
}
\definecolor{skyblue}{RGB}{25, 189, 255} 
\newcommand{\skybluesquare}[1]{%
    \begin{tikzpicture}[baseline=(char.base)]
        \node[draw=skyblue, fill=skyblue, rectangle, rounded corners=2pt, minimum height=10pt, minimum width=10pt, inner sep=0pt, text=white] (char) {\textbf{#1}};
    \end{tikzpicture}%
}
\definecolor{pink}{RGB}{220, 37, 255} 
\newcommand{\pinksquare}[1]{%
    \begin{tikzpicture}[baseline=(char.base)]
        \node[draw=pink, fill=pink, rectangle, rounded corners=2pt, minimum height=10pt, minimum width=10pt, inner sep=0pt, text=white] (char) {\textbf{#1}};
    \end{tikzpicture}%
}
\definecolor{darkgrey}{RGB}{85, 85, 85} 
\newcommand{\darkgreysquare}[1]{%
    \begin{tikzpicture}[baseline=(char.base)]
        \node[draw=darkgrey, fill=darkgrey, rectangle, rounded corners=2pt, minimum height=10pt, minimum width=10pt, inner sep=0pt, text=white] (char) {\textbf{#1}};
    \end{tikzpicture}%
}

\newcommand{\tool}[1]{\texttt{VeriPlan}}
\makeatletter
\def\@ACM@copyright@check@cc{}
\makeatother
\copyrightyear{2025}
\acmYear{2025}
\setcopyright{cc}
\setcctype{by}
\acmConference[CHI '25]{CHI Conference on Human Factors in Computing Systems}{April 26-May 1, 2025}{Yokohama, Japan}
\acmBooktitle{CHI Conference on Human Factors in Computing Systems (CHI '25), April 26-May 1, 2025, Yokohama, Japan}\acmDOI{10.1145/3706598.3714113}
\acmISBN{979-8-4007-1394-1/25/04}
\begin{document}


\title{\texttt{\tool{}}: Integrating Formal Verification and LLMs into End-User Planning}








\author{Christine P Lee}
\orcid{0000-0003-0991-8072}
\affiliation{%
  \institution{Department of Computer Sciences University of Wisconsin--Madison}
  \country{Madison, Wisconsin, USA}
}
\email{cplee5@cs.wisc.edu}

\author{David Porfirio}
\orcid{0000-0001-5383-3266}
\affiliation{%
  \institution{U.S. Naval Research Laboratory}
  \country{Washington, DC, USA}
}
\email{david.j.porfirio2.civ@us.navy.mil}

\author{Xinyu Jessica Wang}
\orcid{0009-0002-5519-8432}
\affiliation{%
  \institution{Department of Computer Sciences University of Wisconsin--Madison}
  \country{Madison, Wisconsin, USA}
}
\email{xwang2775@wisc.edu}

\author{Kevin Zhao}
\orcid{0009-0008-6349-2862}
\affiliation{%
  \institution{Department of Computer Sciences University of Wisconsin--Madison}
  \country{Madison, Wisconsin, USA}
}
\email{kczhao@wisc.edu}

\author{Bilge Mutlu}
\orcid{0000-0002-9456-1495}
\affiliation{%
  \institution{Department of Computer Sciences University of Wisconsin--Madison}
  \country{Madison, Wisconsin, USA}
}
\email{bilge@cs.wisc.edu}
\renewcommand{\shortauthors}{}

\begin{abstract}
Automated planning is traditionally the domain of experts, utilized in fields like manufacturing and healthcare with the aid of expert planning tools. Recent advancements in LLMs have made planning more accessible to everyday users due to their potential to assist users with complex planning tasks. However, LLMs face several application challenges within end-user planning, including consistency, accuracy, and user trust issues. This paper introduces \texttt{\tool{}}, a system that applies formal verification techniques, specifically model checking, to enhance the reliability and flexibility of LLMs for end-user planning. In addition to the LLM planner, \texttt{\tool{}} includes three additional core features---a rule translator, flexibility sliders, and a model checker---that engage users in the verification process. Through a user study ($n=12$), we evaluate \texttt{\tool{}}, demonstrating improvements in the perceived quality, usability, and user satisfaction of LLMs. Our work shows the effective integration of formal verification and user-control features with LLMs for end-user planning tasks.



\end{abstract}





\begin{teaserfigure}
    \includegraphics[width=\textwidth]{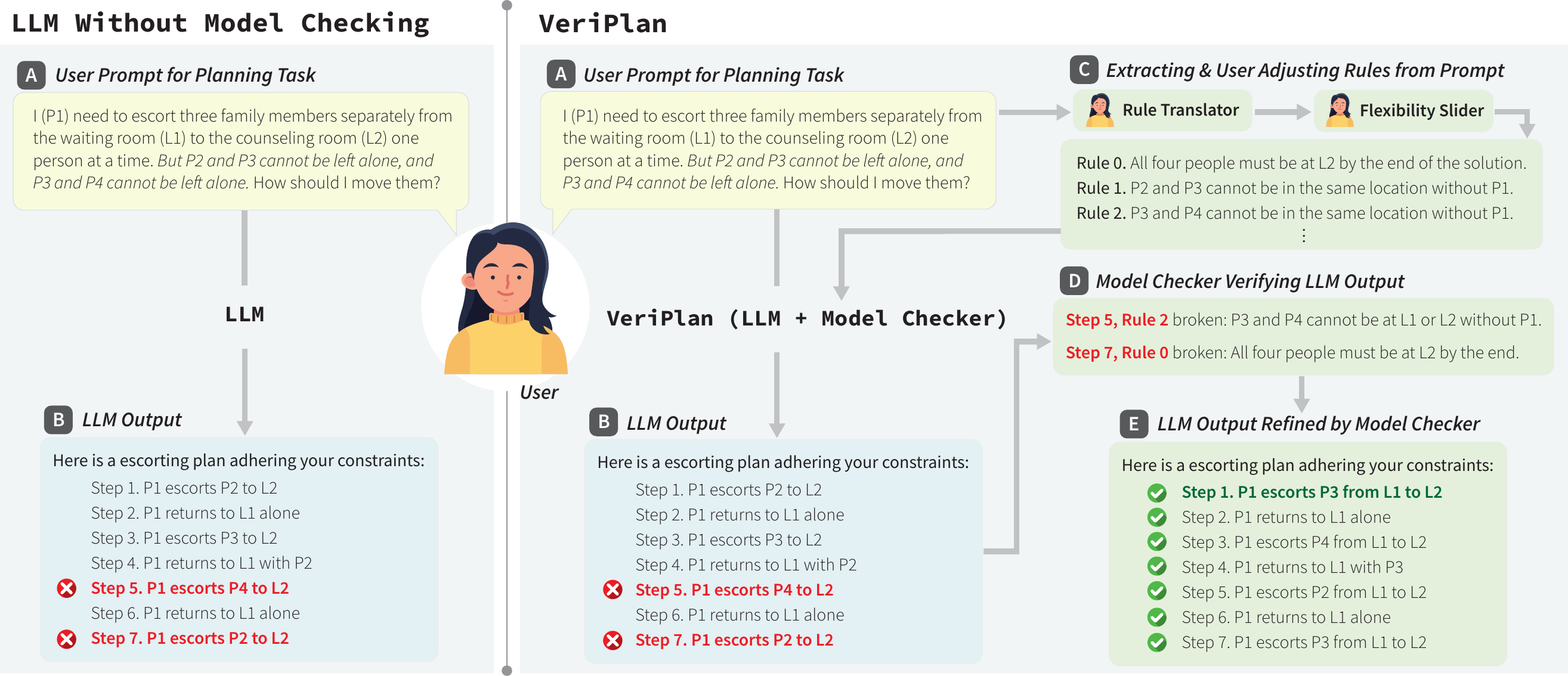}
   \vspace{-12pt}
  \caption{\textit{\ours{} --- }In this work, we present \ours{}, a system that applies formal verification techniques to LLM outputs for end-user planning tasks. The figure compares a user's interaction with an LLM without \ours{} (left) and with \ours{} (right). In both cases, the user provides a prompt requesting a plan with specific requirements (depicted as step \protect\darkgreysquare{A}). The LLM generates an initial planning attempt from the prompt (step \protect\darkgreysquare{B}). With \ours{}, however, rules are automatically extracted from the user's prompt, verified, and refined through direct user involvement (step \protect\darkgreysquare{C}). These rules are then sent to the model checker, which verifies whether the LLM's output adheres to the user-defined rules (step \protect\darkgreysquare{D}). The validation result, along with any rule violations, is shared with both the user and the LLM to refine future planning outputs (step \protect\darkgreysquare{E}). 
  }
  \label{fig:teasor}
\end{teaserfigure}
\maketitle


\section{Introduction}

\textit{Automated planning}---the search for sequences of actions that guide an autonomous agent from an initial state to a goal state \cite{kress2018synthesis}---has traditionally been the domain of experts. Planning has been applied in professional settings, including production planning in manufacturing, medical resource planning in healthcare, project planning in construction, and route and fleet planning in transportation \cite{bourne2011recent,leonetti2016synthesis, lewis2020retrieval}. Automated planning is inherently complex, as the problem space involves managing numerous contingencies, constraints, and variables such as resource limitations, timing dependencies, and evolving preferences or changing conditions. Given the complexity and critical nature of these tasks, entire research communities and industries have dedicated themselves to building and utilizing planning tools (\textit{e.g.,} \cite{fox2003pddl2, nau2021gtpyhop, kapellosaiplan4eu}) that support foresight, decision-making, and the intricate coordination required for effective outcomes.

While these planning tools have traditionally been designed for expert use in professional settings, people increasingly need similar planning support in their everyday lives. People often manage multiple complex planning tasks in their everyday lives, such as coordinating pickup schedules for three children's school and after-school activities, hosting a family dinner party, preparing multiple meals simultaneously, and still setting aside time for personal tasks like writing a book and working out. Despite this need, they often lack effective tools to assist them, relying instead on manual methods or basic calendar apps. 
\revision{Traditional planning tools are often inaccessible for everyday users, as they require expertise in low-level planning languages, complex semantics, or detailed domain specifications for the task environment.}
Recent advances in artificial intelligence (AI), particularly large language models (LLMs), present an opportunity to bridge this gap. By understanding context, adapting flexibly, managing constraints, and automating decision-making, LLMs can make complex planning support more accessible and effective for everyday users.


Despite their potential, end-users have yet to utilize LLMs effectively for such planning tasks. First, it is unclear whether LLMs can, out of the box, offer users solutions that adhere to user expectations, especially in highly constrained planning problems. Existing work has shown that, despite the increasing attention LLMs are receiving as planning tools, they are insufficient for planning and self-verification, particularly in the planning domain \cite{pallagani2024prospects, wang2023describe, huang2024understanding, valmeekam2023planning, kambhampati2024llms, valmeekam2022large}. Recent research has also highlighted several challenges, including difficulties with prompt input and navigation, limitations of text-only interfaces, and issues with evaluating LLMs' consistency and accuracy in meeting user needs \cite{tankelevitch2024metacognitive, subramonyam2024bridging, khurana2024and, sarkar2023exploring}. Finally, LLMs are prone to ``hallucinations''---coherent but incorrect information---that undermine user trust, usability, and satisfaction \cite{leiser2023chatgpt}. These technical limitations and user-centered barriers make it difficult for end-users to rely on LLMs for effective planning.

To address these challenges and enable the effective use of LLMs in end-user planning, LLM-based planning systems must not only be designed to be \textit{reliable}, but the user-LLM interaction must also be designed to support \textit{correction} when the system produces incorrect or unacceptable output---a core principle of human-centered AI \cite{amershi2019guidelines, lee2024rex}. To these ends, LLM-based planning systems must be designed to be (1) verifiable and (2) to keep the user in the loop during verification. Achieving these design principles necessitates combining interaction design with \textit{formal verification}, a set of techniques grounded in mathematical and logical principles to ensure that a system's behavior meets predefined specifications.

In this work, we apply formal verification to LLMs, in order to enable their use as effective end-user planning tools. Specifically, we leverage \textit{model checking}, a formal verification technique, to verify LLM outputs against user-defined constraints. Crucially, we explore how to involve users in the verification process and support user control and flexible adaptation to their needs. Based on this goal, we pose the following research questions: 
\begin{enumerate}
    \item How can formal verification methods, specifically model checking, be effectively applied to LLMs?
    \item How can we engage humans in the process of model checking to improve (1) the quality of outputs from LLMs and (2) the user's experience?
    \item At what stage of the model-checking process should users be engaged to maximize the effectiveness of integrating verification approaches with LLMs?
\end{enumerate}
To address these research questions, we present \tool{}, which integrates a formal verification-based approach to verifying plans generated by LLMs. \tool{} consists of three key features: a rule translator, flexibility sliders, and a model checker, which enables user control throughout the verification process. To evaluate our system, we conducted a user study that ablates different features to assess its effectiveness and impact on users. Our findings indicate that model checking improves the user experience with LLMs in planning tasks, particularly in terms of perceived output quality, user control, and transparency. Additionally, user control over constraint verification enforces rigidity in LLMs, while control over the strictness of constraints enables flexibility and creativity in planning. Finally, we offer design implications for integrating verification methods and user control features into LLM design to make them more useful and applicable for everyday planning tasks.
Our work makes the following contributions: 
\begin{enumerate}
    \item \textit{System contributions:} We present \tool{}, a verification-based approach involving the use of model checking against LLM outputs with multiple user control features. \tool{} includes three key features: rule translator, flexibility adjuster, and model checker. 
    \item \textit{Empirical contributions:} We evaluate \tool{} through a user study $(n=12)$ to understand its effectiveness and the specific contributions of its key features.
    \item \textit{Conceptual contributions:} We present a template-based approach to categorizing temporal constraints for verifying LLM outputs, instantiated and validated within a finite set of scenarios.
    \item \textit{Design implications:} Based on our findings, we present design insights on how to integrate formal verification techniques and user control into the design of LLMs for effective application for end-user planning.
\end{enumerate}

    

\section{Related Works}

\revision{In this section, we provide background on automated planning for end-users and discuss the challenges they face when using LLMs. Next, we review existing verification approaches for LLMs, both broadly and within the context of automated planning. Finally, we provide background on model checking and its use in our verification approach.}

\subsection{Automated Planning for End-users}
\textit{Automated planning} refers to automated techniques that decide \textit{what} an agent does, namely the steps that it takes to achieve a goal, rather than \textit{how} it performs each step \cite{ghallab2016automated}.
Numerous languages and libraries exist that enable users to interact with planning algorithms, such as the \textit{Planning Domain Definition Language} (PDDL) \cite{fox2003pddl2}, the \textit{GTPyhop} planner \cite{nau2021gtpyhop}, and the extensive \textit{Unified Planning} library \cite{kapellosaiplan4eu}, to name a few examples. Although planning tools are typically intended for expert users, recent work has engaged novice users in the planning process through visualization \cite{DePellegrin_Petrick_2024} and plan creation \cite{porfirio2024polaris}.
\revision{However, these planning tools pose significant challenges for end-users due to their reliance on complex formal languages and abstract logic formulas \cite{ebbinghaus1994mathematical, schoen2020authr, hurnaus2010programming}, which are difficult to learn and apply. The technical interfaces often lack intuitiveness, providing rigid workflows and low-level feedback  \cite{helmert2009concise, peer2004pddl, shah2013knowledge}. Moreover, users must invest significant effort in creating detailed system models, specifying states, transitions, and probabilities \cite{porfirio2018authoring, sauer2022structure, porfirio2020transforming}---tasks that demand technical expertise and are highly time-consuming. Designed with a focus on theoretical rigor and correctness, these tools often neglect practical usability, leaving them to fall short in addressing the dynamic and high-level goals of end-users.}

LLMs possess great potential to further increase the accessibility of automated planning for novice users.
Given a natural language prompt or set of prompts, LLMs are demonstrably capable planners \cite{silverLLM2024, songLLMPlan2023, lu2023plug} without requiring the user to directly interact with low-level planning languages or libraries. Still, LLMs are insufficient as standalone planners, requiring external support to verify and improve planning output \cite{pmlr-v235-kambhampati24a}.
To this end, \citet{gundawar2024robustplanningllmmoduloframework} contributes an \textit{LLM-Modulo Framework} that checks LLM-produced plans against a set of \textit{critics}, which provide feedback to the LLM to iterate. In our work, we envision the novice user as a critical component of the verification-feedback loop, akin to recent work in human-LLM interaction for text annotation tasks \cite{wang2024LLM}. For planning tasks, there is a research gap on designing systems to engage novice users in the verification-replanning process, which this work aims to address.

\subsection{End-user Challenges with LLMs}
\revision{
As LLMs are increasingly deployed in everyday applications and engage directly with end-users, they demonstrate great potential but also present significant human-centered challenges, particularly in terms of \textit{usability} and \textit{reliability}.

\textit{Usability} remains a critical issue as users frequently struggle with crafting effective prompts and engaging with systems beyond the input stage. Studies highlight the difficulty users face in formulating prompts that elicit desired responses \cite{zamfirescu2023johnny, khurana2024and, tankelevitch2024metacognitive, subramonyam2024bridging, liu2024we}. Additionally, the cognitive demands placed on users---such as monitoring and deciding on strategies for prompting and interaction---exacerbate these challenges \cite{tankelevitch2024metacognitive, subramonyam2024bridging}. 
Another usability barrier is users' difficulty understanding how prompts influence outputs and building accurate mental models of the system's behavior and the reasoning behind it \cite{bhatt2021uncertainty, sun2022investigating, vasconcelos2023generation}. 
In response to these challenges, engaging users during the interaction process to steer the LLM's behavior, and support user's understanding of the reasoning has gained increasing attention. 
Strategies like co-creation, where users and AI collaboratively refine outputs, have been proposed to expand engagement and improve interaction intuitiveness \cite{schellaert2023your}. Similarly, interactive environments with user-controllable parameters enable experimentation, helping users build a better understanding of LLM capabilities \cite{louie2020novice, ma2024beyond, suh2023sensecape, jiang2023graphologue}. 
In addition, approaches like enhancing explainability and introducing customizable interaction options aim to reduce cognitive load and improve user experience \cite{tankelevitch2024metacognitive, teufelberger2024llm}. 
While engaging users and providing control to address usability challenges is a promising direction, further work is needed to understand \textit{how} and \textit{when} to involve users throughout the interaction process with LLMs. Such exploration can reveal ways to gather direct input and feedback that help LLMs accommodate evolving preferences and more effectively meet diverse user needs.

The \textit{reliability} of the output is another significant challenge. LLMs are prone to generating text that appears structurally coherent but contains factual inaccuracies or nonsensical information, a phenomenon known as hallucination \cite{rawte2023survey, bender2021dangers, ji2023survey, maynez2020faithfulness}. The lack of interpretability further complicates users' safe reliability, as users often struggle to understand the reasoning behind the output of the LLM \cite{mathews2019explainable, zhao2024explainability, yang2024harnessing, mirchandani2023large, liu2024we}. These issues are especially concerning in safety or mission-critical domains, such as healthcare or military applications, where reliance on incorrect outputs can have severe consequences \cite{koga2023exploring, lee2023benefits, sallam2023chatgpt}. These issues can further lead to risks of users over-relying on LLM-generated outputs without sufficient critical evaluation, underscoring the need for mechanisms that support users' safe and reliable interactions with LLMs \cite{ji2023survey, maynez2020faithfulness}. 
}

\subsection{Verification Approaches for LLMs}
\revision{
The advancements in LLMs have unlocked unprecedented capabilities in sense-making, language use, and interaction, enabling precise inference of user needs and applications across diverse domains \cite{kim2024understanding, minaee2024large, zhao2023survey}.
As these systems advance, ensuring their safety, reliability, trustworthiness, and alignment with user needs has become a pressing focus. To address this, a substantial body of work has emerged on verifying LLM outputs, which we broadly categorize into two directions.

The first direction focuses on enhancing user trust through explanations and interface design. Existing approaches generate explanations to support users in understanding and trusting LLM outputs \cite{li2022explanations, krishna2024post, huang2023can, marasovic2021few, wiegreffe2021reframing}. Others have explored designing interfaces and tools that help users deconstruct textual components, evaluate LLM outputs, and act upon them effectively \cite{ma2024beyond, suh2023sensecape, jiang2023graphologue}.

The second direction focuses on ensuring the validity of LLM outputs. One notable direction includes using LLMs for evaluation \cite{desmond2024evalullm, zhang2023wider, zheng2024judging} or orchestrating multi-agent systems to verify outputs \cite{mostajabdaveh2024optimization, liang2024improving, hassan2024llm, chan2023chateval}. These methods have been applied to complex tasks such as mathematical reasoning \cite{wu2024mathchat, zhang2025mathverse, li2023making}, semantic reasoning \cite{chen2023teaching, ni2023lever, liu2024speak}, and data annotation \cite{wang2024human}. Additionally, other approaches involve humans in evaluating and correcting outputs \cite{wang2024human, shankar2024validates}. Finally, a growing area of research incorporates constraint-based approaches, such as applying constraints to planning in robotics \cite{yang2024plug}, creating datasets with constraints for evaluation \cite{zhang2024cfbench}, or generating plans that adhere to multiple constraints \cite{xie2024travelplanner}. However, constraint-based approaches often utilize predefined datasets and can suffer from the lack of mechanisms for dynamically incorporating user preferences, needs, or evolving contexts.

Despite recent advancements, challenges persist in relying on LLMs for verification. Using LLMs to verify their own outputs risks critical flaws. Studies highlight their deficiencies in error detection, correction mechanisms, and adherence to constraints, as well as their tendency to hallucinate or retrieve inaccurate context \cite{ji2024testing, yao2023llm, liu2024exploring}. 
For instance, in the planning domain, despite extended context windows and few-shot learning, \citet{xie2024travelplanner} and \citet{chen2024can} demonstrate that LLMs struggle to generate plans and feedback for complex scenarios or adhere to predefined constraints. Similarly, \citet{valmeekam2023planning} reports that GPT-4 achieves an average success rate of 12\% in planning tasks, highlighting the inadequacy of LLMs in handling intricate requirements independently.
Other works have highlighted how utilizing LLMs for evaluation can suffer from bias based on the order, appearance, or length of the content, aspect-specific evaluation, scalability, and effectiveness in diverse contexts \cite{wang2023large, huang2024empirical, koo2023benchmarking, son2024llm, park2024offsetbias}. 
These limitations have led to heuristic and modular approaches as verification mechanisms to address such shortcomings \cite{kambhampati2024llms, valmeekam2022large}. Moreover, LLM reasoning and explanations, such as chain-of-thought reasoning, can be influenced by biased contexts, raising further caution about their reliability \cite{turpin2024language}. Consequently, developing methods to verify LLM outputs without relying on LLMs is critical to ensure validity, particularly for high-stakes, real-world applications.

}



\subsection{Model Checking in Formal Verification and LTL constraints}\label{hello}

Model checking is a formal verification technique used to determine whether a software or hardware system satisfies requirements expressed in formal logic \cite{baier2008principles}. By systematically exploring all possible states that a system may encounter or produce, model checking exhaustively examines system behavior against these requirements, making it essential for proving the behavior of highly complex systems. Linear Temporal Logic (LTL) is a commonly used representation to express requirements, or \textit{properties}, in domains such as assistive robotics \cite{dixon2014fridge} and autonomous navigation \cite{liu2023grounding}. LTL allows users to specify and compose temporal constraints in the form of sequencing (\ie{} ``event A must occur before event B''), eventuality (\ie{} ``event C must eventually happen.''), and safety (\ie{} ``event D will never occur''), to name a few examples. This expressiveness makes LTL suitable for real-world tasks such as scheduling, safety protocols, and workflow management, where the timing and the order of actions are critical. 

\revision{In summary, our work builds on existing approaches to verify and validate LLM outputs, with a particular focus on constraint-based methods. We extend these methods by directly involving human engagement to define and refine constraints that align with users' needs and preferences. Our features for human engagement are designed to support varying levels of user control and involvement, for users to effectively guide the LLM's behavior. We leverage the significant potential of LLMs as end-user planning tools while addressing their shortcomings and user challenges through the implementation of an external verification approach using model checking, a formal verification technique.}

\begin{figure*}[!th]
  \includegraphics[width=\textwidth]{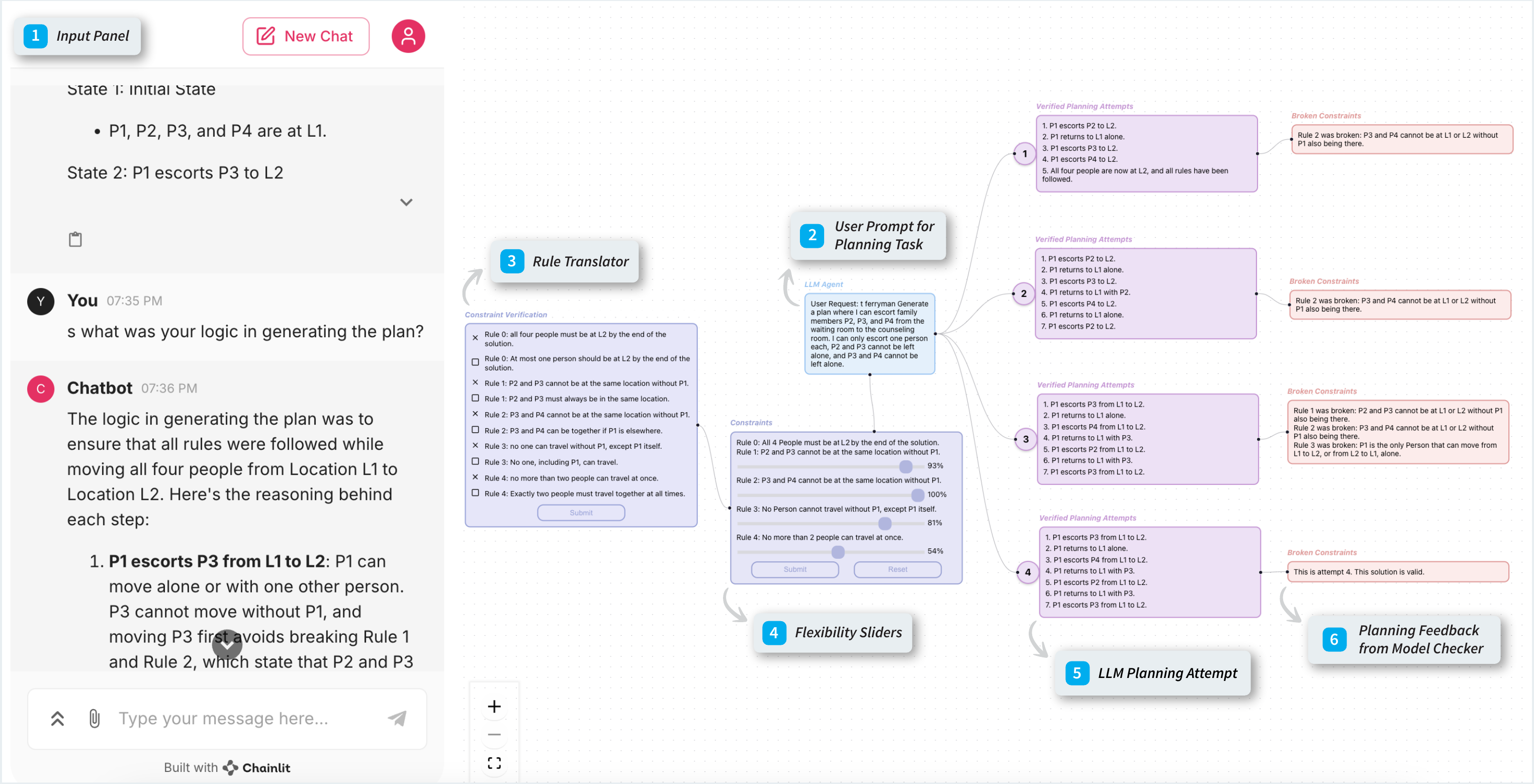}
   \vspace{-12pt}
  \caption{\textit{\ours{} Interface ---} The front-end interface of \ours{}. We outline the user's interaction with the front-end as a guide to explain the pipeline of \ours{} in Section \ref{sec:tech}.
}
  \label{fig:interface}
\end{figure*}

  


\section{Technical Approach}\label{sec:tech}
This section introduces the technical approach of \ours, illustrating how it utilizes model checking on LLM outputs. We begin by outlining the three core features of \ours, followed by a detailed explanation of the technical approach for each feature, accompanied by an illustrative example. All LLM agents used in our implementation are powered by GPT-4 \cite{gpt-4}. Specific information on prompts used for LLM agents and the source code for our implementation can be found in the supplementary materials.\footnote{The supplementary materials can be found at \url{https://osf.io/va6d5/?view_only=8d74c81f765746908420e63479f6f36d}} 

\subsection{Patient Navigation Planning Scenario} \label{sec:scenario}

Throughout this section, we use the scenario of a user using an LLM to plan patient navigation for a counseling session while following conflict-prevention rules to illustrate how \ours{} assists with complex planning tasks.

\begin{quote}
    \textit{You (P1) are a family counselor preparing to hold a family therapy session. You are aware that certain family members have deeper conflicts with some more than others. You believe that a group session could be beneficial, allowing you to use established procedures to help heal family tensions. However, to avoid conflict before the group session begins, you decide to escort each member separately to the counseling room (L2) based on the severity of their conflicts. All family members are currently in the waiting room (L1) with you. Due to hospital safety protocols, all family members (P2, P3, P4) must be escorted by you, and only one person can be escorted at a time. However, because of ongoing tensions, P2 and P3 cannot be left alone together, and similarly, P3 and P4 cannot be left alone together.}
\end{quote}

Using this scenario, we demonstrate how \ours{} assists the user in iteratively solving the navigation planning task using its three features for model checking---rule translator, flexibility slider, and model checker---until a successful planning solution is reached.

\subsection{Features}
The verification approach implemented in \ours{} includes the following features: the (1) LLM planner; (2) rule translator; (3) flexibility slider; (4) model checker; and (5) refined LLM planner. 
\paragraph{LLM Planner}
The LLM receives the initial user input in the form of a natural language prompt, which includes the user's request, context, and constraints. Based on this input, the single-agent LLM will attempt to create a plan according to the provided prompt.

\paragraph{Rule Translator}
The rule translator converts the user's initial natural language input into formal language properties that are interpretable for the model checker to use during verification. 
The translation is then translated back into natural language and presented to the user, who provides feedback to verify whether the translation is accurate.

\paragraph{Flexibility Slider}\label{hello}
Once the correctness of the rules is verified, the user can adjust the strictness of each rule using the flexibility sliders, defining the level of enforcement. This strictness determines the extent to which the model checker will insist on adhering to the rules during model checking. 

\paragraph{Model Checker}
\revision{\ours{} employs an external verification process, using a formal verification technique called model checking (see \S\ref{hello} for more). For model checking, we use an off-the-shelf probabilistic model checker, to systematically inspect every state within the system to confirm whether a set of behavioral properties are satisfied.} The model checker uses the user-defined constraints to evaluate the LLM planner's planning attempts, ensuring they align with the specified requirements. After completing the evaluation, the model checker provides feedback to the user and LLM on whether the plan is valid or which constraints are violated.

\paragraph{Refined LLM Planner}
Once feedback is provided, the LLM planner will iteratively regenerate a plan based on this feedback until it either reaches a valid solution or the maximum number of iterations specified in the program. At the end of the iterations, based on feedback from the model checker, the user can adjust the constraints using the rule translator or flexibility sliders before rerunning the LLM planner to reach a satisfying solution.



\begin{figure}[!h]
  \includegraphics[width=\columnwidth]{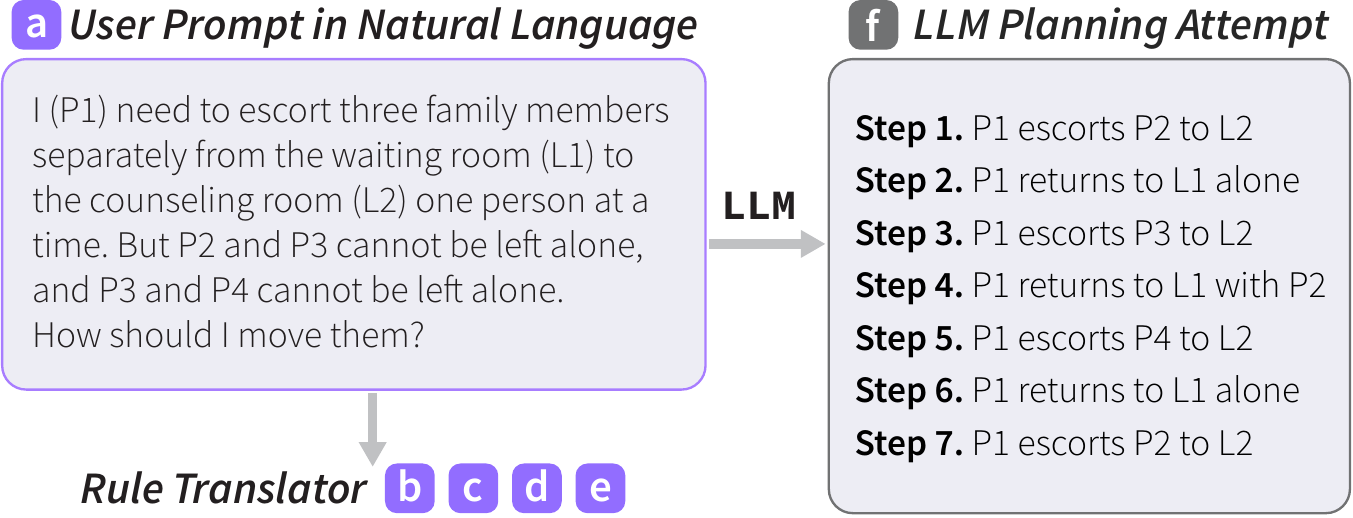}
   \vspace{-12pt}
  \caption{\textit{\ours{} LLM Planner ---} 
  Pipeline of the LLM planner described in Section \ref{sec:llmPlanner}. When the user submits an initial prompt in natural language, an LLM agent generates a plan based on the user's input. This plan is later to be verified by the model checker. Simultaneously, the user's prompt is sent to the rule translator to initiate the verification process. 
}
  \label{fig:LLMplanner}
\end{figure}

\subsection{LLM Planner}\label{sec:llmPlanner}
The front-end interface of \ours{} is shown in Figure \ref{fig:interface}.
In the example scenario, the user inputs their full planning requests and constraints through the input panel (depicted as step \skybluesquare{1}), and the request is reflected on the interface (step \skybluesquare{2}). 

\subsubsection{How It Works}
The beginning of the pipeline for \ours{}, including the LLM planner, is presented in Figure \ref{fig:LLMplanner}.
At the start of the interaction, as the user inputs their prompt (step \purplesquare{a}), an LLM agent generates an initial plan based on the user's request (step \darkgreysquare{f}). This plan is then later to be checked by the model checker, using the constraints defined by the rule translator and the flexible slider features discussed below.

\begin{figure*}[!b]
  \includegraphics[width=\textwidth]{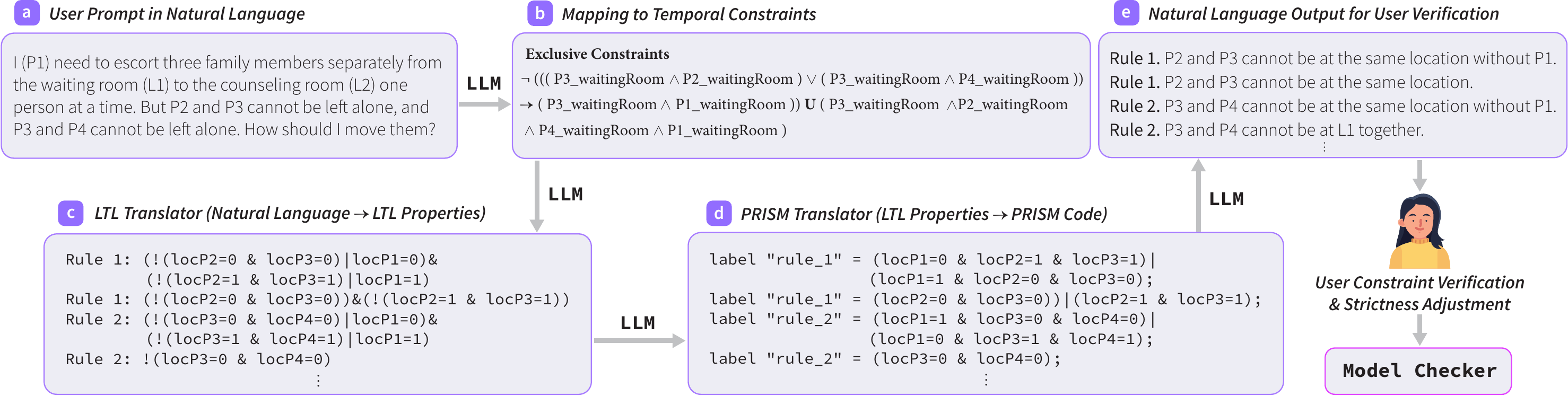}
   \vspace{-12pt}
  \caption{\textit{\ours{} Rule Translator ---} Pipeline of the rule translator described in Section \ref{sec:ruleTranslator}. The translator extracts a set of constraints from the user's initial natural language input that must be adhered to for a correct plan. These constraints are mapped to appropriate LTL properties within the temporal constraint template (Table \ref{fig:template}) for model checking. Using this template, the constraints are converted into LTL and PRISM language for model checking, and then presented back in natural language for user verification.
}
  \label{fig:pipeline}
\end{figure*}

\subsection{Rule Translator}\label{sec:ruleTranslator}

The role of the rule translator is to extract constraints from the user's prompt that a correct plan must follow.
\revision{The rule translator presents the extracted results to the user, allowing them to review the extracted constraints and either confirm them or request regeneration. For confirmation, the user selects the correct version of the constraint using the check box (step \skybluesquare{3}).} If the presented constraints are unsatisfactory, the user can ask \revision{the rule translator} to regenerate translations for the constraints using the input panel.

\subsubsection{How It Works}
The pipeline of the rule translator is shown in Figure \ref{fig:pipeline}.
Receiving the user prompt (step \purplesquare{a}) which includes the user's planning request and desired constraints, \revision{an LLM-based mapping agent extracts content from the prompt and maps it to the appropriate categories in the} \textit{temporal constraint template} described below (step \purplesquare{b}). The mapping agent is bound to select from the seven categories and has been prompt-engineered with examples for mapping accuracy.

\begin{table*}[!th]
  \includegraphics[width=\textwidth]{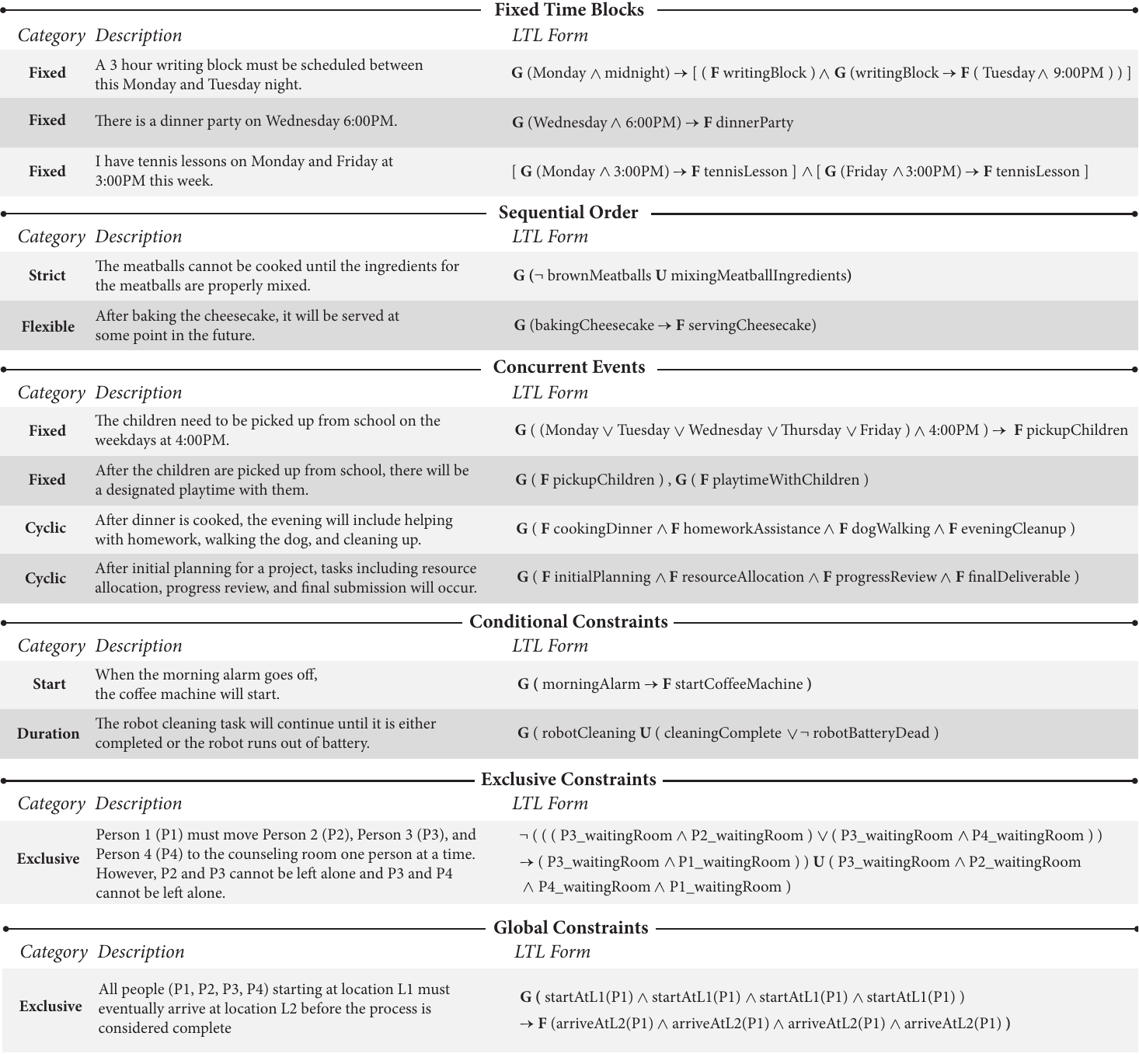}
   \vspace{-12pt}
  \caption{\textit{Template of Temporal Constraints ---}  List of temporal constraints used by the Rule Translator (Section \ref{sec:ruleTranslator}) that translates natural language into LTL properties. These constraints were instantiated and validated within a finite set of scenarios described in Section \ref{sec:scenarios}.
}
  \label{fig:template}
\end{table*}

\paragraph{Template of Temporal Constraints}
To ensure that the rule translator can accurately convert user input into rules for model checking, it uses a predefined temporal constraint template. 
For the model checker to function, the rules must be specified in LTL logic. However, since users input rules in natural language, manually translating them into LTL formulas is challenging. \revision{Unlike fixed algorithms that require rigid input formats, LLMs can interpret and categorize variable natural language inputs into temporal categories by understanding context and intent, guided by examples from prompt engineering. This adaptability allows complex or unconventional rules to be mapped to predefined LTL constraint templates, reducing the need for extensive manual refinement in rule translation.}

To address this, we developed a template of LTL properties which are fed into an LLM for translation, covering six temporal categories: (1) fixed time blocks, (2) sequential order, (3) concurrent events, (4) conditional constraints, (5) exclusive constraints, and (6) global constraints. Each category includes a template for converting natural language into LTL properties, which are fed into the LLM. 
\revision{In the constraint template, LTL provides \textit{modal operators} to formalize such statements. The global operator, $G$, specifies conditions that must hold in every state. The future operator, $F$, checks for events that must occur at some point in the future. The until operator, $U$, specifies that an event $\phi$ must remain true until another specified event $\psi$ occurs, and that $\psi$ must indeed happen.}
The detailed templates are provided in \revision{Table \ref{fig:template}}.


Once the mapping is complete, it is sent to the LTL translator (step \purplesquare{c}). The LLM-based LTL translator uses the template to convert the mapped outputs into LTL properties, guided by prompt engineering to determine the appropriate conditionals for each constraint. The translator then generates an LTL formula for the constraint.


These LTL translations are then sent to the LLM-based PRISM translator, for converting the LTL properties into an interpretable format for the model checker (step \purplesquare{d}). Our verification approach utilizes the PRISM Model Checker \cite{kwiatkowska2011prism} (discussed in detail in \S\ref{sec:modelChecker}) to format LTL properties, which requires that properties be expressed in the PRISM language. \revisionNew{While an algorithmic approach could perform this translation, an LLM was chosen for its seamless integration and demonstrated feasibility during system design.} Our PRISM translator utilizes manual examples for prompt engineering to convert LTL expressions into the PRISM format, covering state representations, rule violations, and temporal logic translations.


These two sets of translations are then sent to the user for final verification and confirmation of each constraint. Before being presented to the user, each translation is converted back into natural language by the PRISM and LTL translator for user readability (step \purplesquare{e}).
\revision{The translated rules are presented to the user in natural language for review. The user can verify their correctness and make adjustments if needed. If a rule aligns with the user's expectations and goals, the user can confirm it by marking the checkbox next to it; otherwise, they can provide feedback to regenerate the rule using the rule translator. Only the rules with marked checkboxes will be included in the final set. Once all desired rules are confirmed, the user finalizes the process by selecting the `submit' button.}
Based on user input, the final set of rules to be used for model checking is finalized along with the corresponding LTL properties and PRISM code. The final set of constraints is then passed to the flexibility sliders for strictness adjustment.

\subsection{Flexibility Sliders}
As shown in Figure \ref{fig:interface}, once users have verified the correctness of the constraints, they can specify the strictness of each constraint using the flexibility sliders (step \skybluesquare{4}). In the given example, the user initially believes that all the rules should be treated as hard constraints, as they pertain to hospital protocols and are crucial for avoiding conflicts among patients. Consequently, they set the sliders to 100\% for each rule and submitted the adjustments. After the first few attempts fail, the user decides to set the strictness of rule four to 50\%, reasoning that P1 might be able to travel with both P2 and P4. Throughout the interaction, users can freely modify the strictness of individual rules after reviewing the outputs from the LLM and model checker. Once the strictness levels are finalized, the complete set of constraints, verified and customized by the user, is sent to the model checker.

\subsubsection{How It Works}
Constraints that are verified by the user from the rule translator are then sent to the flexibility sliders. These sliders allow users to adjust the strictness of each rule, where strictness defines how rigidly the model checker will enforce the rule. Strictness includes both ``soft'' and ``hard'' constraints: hard constraints \textit{must} be satisfied for a plan to be valid, and any plan that violates a hard constraint is immediately rejected. Soft constraints, while preferred, are not strictly necessary and their violation does not invalidate the plan. \revision{If a soft constraint is violated, unlike hard constraints, the plan will not be immediately rejected. Instead, the plan with the violated soft constraint will be marked as valid, and the user will be notified of the violation.}
Constraints are then weighted based on hardness, and \ours{} samples from the weighted constraints, with lower-weighted constraints (corresponding to ``softer'' constraints) being less likely to be sampled. The model checker then checks the plan against the sampled constraints.


\begin{figure*}[!b]
  \includegraphics[width=\textwidth]{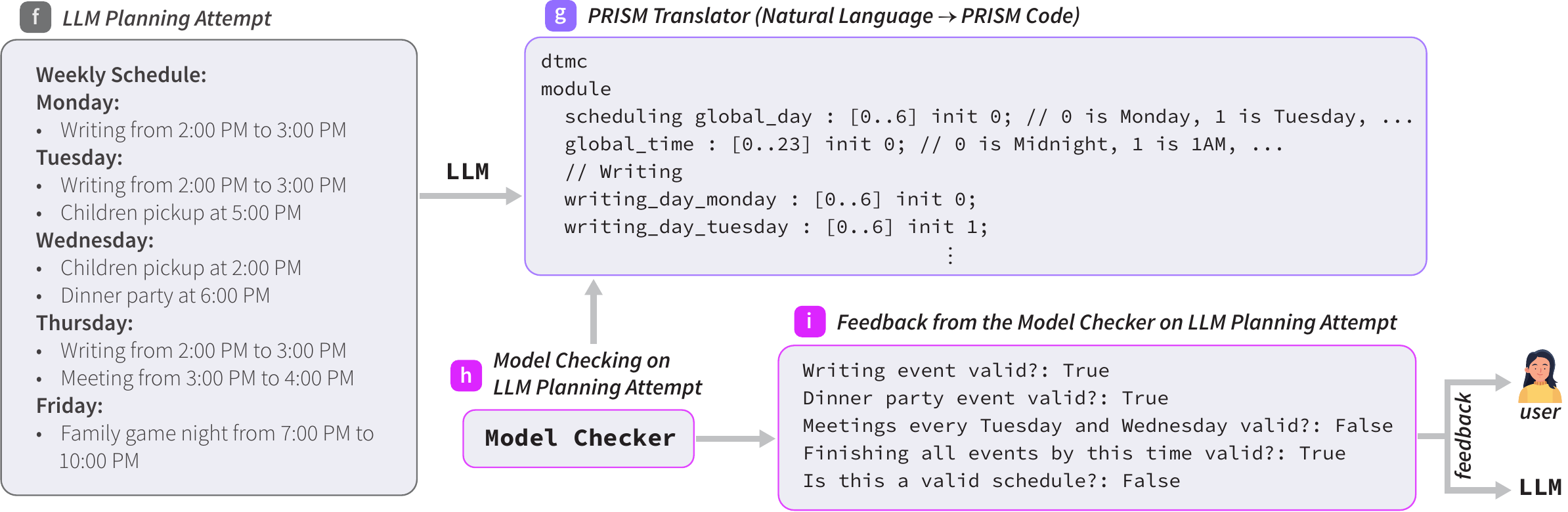}
   \vspace{-12pt}
  \caption{\textit{\ours{} Model Checker ---} Pipeline of the model checker described in Section \ref{sec:modelChecker}. The model checker takes in the LLM-generated plan in PRISM language and the set of rules from the previous stage, and then evaluates the plan against these rules. The plan's validity, along with any violated rules, is sent to both the user and the LLM agents to refine future solutions.
}
  \label{fig:modelChecker}
\end{figure*}

\subsection{Model Checker}\label{sec:modelChecker}
Once the correctness and strictness of the rules are defined by the user, the model checker uses these rules to check the initial plan generated by the LLM agent. In the interface, the user can view the initial planning attempt generated by the LLM (step \skybluesquare{5}). The model checker then performs model checking on this plan, comparing each state against the specified constraints. Based on the verification, the model checker provides feedback, which includes a list of broken rules or confirmation of the plan's validity (step \skybluesquare{6}). This feedback is then sent to the user to explain the system's status and to the LLM for regenerating the plan based on the feedback. 

\subsubsection{How It Works}
The pipeline of the model checker is shown in Figure \ref{fig:modelChecker}. 
Similar to LTL translation, the initial plan generated by the LLM agent based on the user's request (depicted as step \darkgreysquare{f}) is also translated into the PRISM language format for the model checker to process (step \purplesquare{g}). In this work, we use the PRISM Model Checker \cite{kwiatkowska2011prism} and Stormpy for verification. Stormpy is a Python API for Storm \cite{hensel2022probabilistic} that enables model checking and property verification within a Python environment. At this point, since the model checker has (a) a set of LTL-expressed rules, and (b) the LLM-generated plan expressed in the PRISM language, it evaluates the plan against these rules (step \pinksquare{h}). During verification, the model checker examines each state of the plan for rule violations. Any rule violations will result in an invalid plan. The validity of the plan, along with any rules that were violated are sent to both the user and the LLM agents to refine their future solutions (step \pinksquare{i}).


\subsection{Refined LLM Planner}
Receiving the feedback from the model checker, the process of the LLM regenerating a plan and the model checker verifying it against the user-defined rules is iterated two additional times, allowing for a total of three iterations, as defined by the system parameters.
Between iterations, the user can adjust the strictness of the constraints to explore different planning solutions (step \skybluesquare{4}). 
Once all iterations are complete, the user can choose to inquire about aspects such as the constraints, the decision-making procedure, the logic of the model checker, or the system status through the input panel (step \skybluesquare{1}). Additionally, the user can modify the constraints using the flexibility sliders (step \skybluesquare{4}), or modify the constraints through the rule translator through the input panel before initiating a new interaction (step \skybluesquare{1}).

\subsubsection{How It Works}
Upon receiving feedback from the model checker, this information is provided as updated requirements to the LLM, which is then asked to regenerate a plan. The regenerated plan is checked by the model checker for rule violations using the user-defined rules. If no violations are found and a correct plan is generated, the interaction ends. If a correct plan is not generated by the end of the iterations, the system prompts the user to adjust the constraints or their strictness for additional iterations.
\section{User Study}
\subsection{Scenarios}\label{sec:scenarios}
We design three scenarios that incorporate the temporal constraints illustrated in Table \ref{fig:template}. One of these scenarios is the ``patient navigation in hospital'' example discussed in \S\ref{sec:scenario}. Below, we describe the remaining two scenarios.

\paragraph{Optimizing Cooking Procedures}
The user is hosting a dinner party on Wednesday at 6:00 PM with multiple guests, requiring the preparation of various dishes to accommodate different dietary preferences, such as vegetarian and gluten-free. The user plans to make spaghetti and meatballs as the main dish and cheesecake for dessert, with meat, vegetarian, and gluten-free versions of each. The user must plan how to cook these dishes simultaneously, ensuring they are ready on time while optimizing the cooking process.

\paragraph{Scheduling Multiple Events}
The user is trying to schedule multiple events for the week. These include three hour writing blocks for her book, a dinner party on Wednesday at 6:00 PM, meetings with colleagues on Tuesdays and Wednesdays, tennis lessons on Fridays at 3:00 PM, child pickup and playtime, household chores, and personal routines (\textit{e.g.,} listening to music while writing or having coffee in the morning). Every Sunday evening, she creates a weekly plan to organize and fit all these tasks into her schedule.

\subsection{Study Design}
This study aimed to understand the importance and impact of \ours{}'s verification approach and user control features, specifically evaluating how these elements influenced user reliance, usability, satisfaction, and the perceived performance of LLM outputs. 
We conducted an ablation study using a within-subjects design, where different ablation conditions served as the within-subjects variable.
In Condition 1, participants engaged with \ours, which included the rule translator, flexibility sliders, and model checker. Condition 2 removed the flexibility slider, leaving only the rule translator and model checker. Condition 3 removed the rule translator, including only the flexibility slider and model checker. In Condition 4, all three features, including the rule translator, flexibility slider, and model checker, were removed as neither the rule translator nor the flexibility sliders can function without the model checker. For consistency, we denote these conditions with \textit{C1 (Full)}, \textit{C2 ($\neg$Slider)}, \textit{C3 ($\neg$Translator)}, \textit{C4 (None)} in the remainder of the paper.
\revision{During the study, participants were randomly assigned to two of the three scenarios. In each scenario, participants engaged in all four conditions in a randomized order. After each condition, participants completed the quantitative scales. At the end of their interaction with each scenario, semi-structured interviews were conducted.} The entire study lasted 1.5 hours. Questionnaires used during the study can be found in the supplementary materials.\footnote{The supplementary materials can be found at \url{https://osf.io/va6d5/?view_only=8d74c81f765746908420e63479f6f36d}} 

\subsection{Measures}
To evaluate the participants' experiences with the system, we employed the Usefulness, Satisfaction, and Ease (USE) scale \cite{article} to measure three key dimensions: usefulness (Cronbach's $\alpha = 0.94$), ease of use (Cronbach's $\alpha = 0.83$), and satisfaction (Cronbach's $\alpha = 0.95$). We also used the performance questionnaire from the fairness, accountability, transparency, and explainability (FATE) scale developed by \citet{shin2021effects} to measure participants' perceived quality of the LLM's output (Cronbach's $\alpha = 0.91$). Both scales were placed on a seven-point Likert scale.

\subsection{Participants}
12 participants were recruited for our user study. Participants were required to be in the United States, fluent in English, and at least 18 years old. All participants were recruited through university mailing lists. 
\revision{While our sample size is not large, the within-subjects study design achieves an acceptable level of statistical power for significant results \cite{bellemare2014statistical}.} 
Participants age ranged from 19--48 ($M = 25$, $SD = 7.9$). 50\% of the participants identified as female and 50\% as male. 50\% of our participants were White, 41.6\% were Asian, and 8.4\% were American Indian or Alaska Native. After the study, participants were compensated \$15.00 per hour. We refer to participants as P1--P12, using the notation P\textit{i} to indicate participants, where \textit{i} indicates participant ID number.
\revision{In the recruitment survey, we also collected participants' experiences with LLMs, asking them to select a category that best described their familiarity: ``not familiar or none,'' ``occasional use,'' or ``regular use.'' Five participants (P7--P11) selected ``not familiar or none,'' four (P1, P4, P6, P12) selected ``occasional use,'' and three (P2, P3, P5) selected ``regular use.'' Those who reported occasional or regular use mentioned using LLMs for tasks such as brainstorming, search engines, writing assistance, and planning tools (\eg scheduling assistance, task management, project coordination, and itinerary planning.)}

\subsection{Analysis}

\revision{For the quantitative data, we conducted a Dunnett test to compare the means of the ablation groups (C2, C3, C4) to the mean of the full system (C1). Dunnett's test compares the mean of several experimental conditions to a control condition, in which for our study, the full \ours{} system (C1) is considered to be the control. The test was performed with an alpha level of 0.05.}

For qualitative data, we conducted a Thematic Analysis (TA) on the interviews. The coding of the responses was conducted by deriving representative themes from transcriptions~\cite{clarke2014thematic, McDonald19}. During open coding, the first author coded for significant concepts in the data. Concepts were then categorized into clusters, further being grouped into themes. These themes were iteratively discussed between the whole research team, recategorizing the groups and revising the themes upon disagreement until a consensus was reached. 

\begin{figure*}[!th]
  \includegraphics[width=\textwidth]{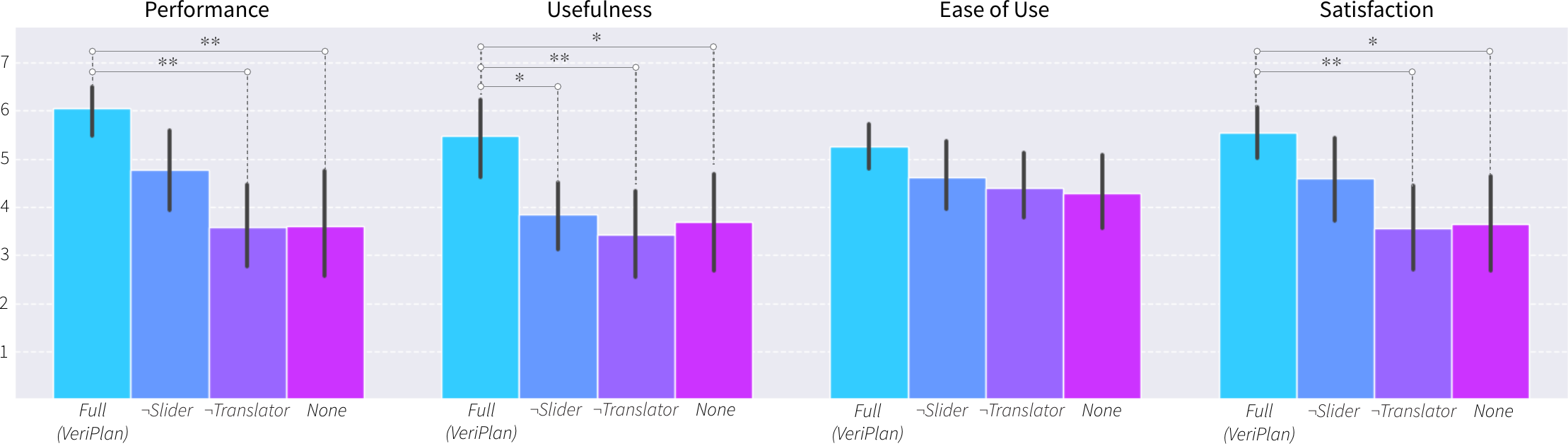}
   \vspace{-12pt}
  \caption{\textit{Quantitative Data from User Study ---} Bar graphs on participants' perceived performance of LLM, usefulness, ease of use, and satisfaction scores across different conditions. Horizontal lines indicate significant pairwise comparisons with \revision{Dunnett's test} ($p < 0.05^{\ast}$, $p < 0.01^{\ast\ast}$, $p < 0.001^{\ast\ast\ast}$). Vertical lines in each bar graph indicate standard error. 
}
  \label{fig:userStudyQuant}
\end{figure*}

\section{Results}
Our analysis aimed to understand the impact of our verification-based approach and its features on the effectiveness of and user experience with LLMs in planning tasks. The results of our quantitative data are shown in Figure~\ref{fig:userStudyQuant}. 
\revision{Overall, the Dunnett test revealed that the full system (C1) significantly outperformed the \textit{C3 ($\neg$Translator)} $(p=.0011)$ and \textit{C4 (None)} $(p=.0013)$ conditions; was significantly more useful than the \textit{C2 ($\neg$Slider)} $(p=.047)$, \textit{C3 ($\neg$Translator)} $(p=.009)$, and \textit{C4 (None)} $(p=.0257)$ conditions; and was significantly more satisfying than the \textit{C3 ($\neg$Translator)} $(p=.007)$ and \textit{C4 (None)} $(p=.0101)$ conditions.}

Below, we present our findings in \revision{four} key themes that emerged in our analysis. For the first three themes, we present quantitative findings first, followed by qualitative insights that reveal differences in use patterns and user perceptions across conditions, providing a deeper understanding of our system's impact.
\revision{For the fourth theme, we present findings derived from the qualitative analysis, focusing on participants' interaction experiences with \ours{}.}

\subsection{Rule Translator Improved Performance, Satisfaction, and Usefulness of LLMs}
Participants' scores in \textit{C3 ($\neg$Translator)} were significantly lower than those in \textit{C1 (Full)} in measures of performance, usefulness, and satisfaction. Our qualitative analysis provides further insight into these results.

\subsubsection{Verification Enabled Users to Control Rigidity}
All twelve participants noted that the ability to verify translations and adjust rules provided a sense of user control. This control allowed them to set deterministic boundaries, creating a level of rigidity within which the LLM could operate when generating plans. This rigidity ensured that the LLM's outputs were aligned with their personalized, user-defined constraints, matching their needs and goals. As P10 explained: \textit{``I know LLMs work probabilistically, so having these rigid boundaries felt like it was getting better accuracy. And because I defined those boundaries, they were useful to me. I liked that I was giving the algorithm more defined limits to create plans that fit me.''}

Seven participants (P2, P5--P8, P10, P11) emphasized that being involved in the verification process was core to ensuring the validity and correctness of the user-specified rules, which directly improved the system's usability and reliability. In contrast, when they were unable to participate in defining the rules, users were less confident that the system accurately reflected their goals or needs, leading to reduced trust in the output. As P6 noted: \textit{``I liked being involved in planning the output, knowing these are my rules, not the computer's. It helps with the validity of what's being spit out. Especially in personal situations, where I want more input opportunities to ensure the system doesn't misinterpret what I want.''}



In Condition 3 where the rule translator was ablated, five participants (P3, P7, P9, P11, P12) decided to stop interacting after at most two attempts because the system failed to correctly translate their prompts into rules, and adjustments were unsuccessful. As P3 explained: \textit{``Because I know the constraints are wrong, I don't want to do anything after this. So then every solution it generates, I'm gonna have to double-check anyway, so it's probably more efficient for me to just use my brain to generate my own solution. I can't verify that the constraints are 100\% correct, so I can't trust the material it produces. It's not very helpful.''}

\subsubsection{Verification Helped Align Expectations and Refine Prompts}

By being involved in the validation of the rules, eight participants (P1, P2, P3, P5, P7--P9, P11) found this procedure helpful in ensuring that the LLM's rules and inputs aligned with their expectations and goals. Through the translation and verification process, participants aimed to understand how the system interpreted their prompts, minimizing gaps or misunderstandings during translation. As one participant explained:
\textit{P8: ``I think it [using the rule translator] was more like fact-checking for reliance and trust, ensuring that it [LLM] is listening to what I'm saying and will actually give me a plan that adheres to my facts.''}

Five participants (P1, P3, P4, P7, P12) also noted that the verification process was effective in mitigating the impact of messy or unorganized prompts. They described their prompts as often being a text dump from their minds, sometimes lacking key details. The system's translation and presentation of prompts as rules helped participants organize their thoughts and check for completeness of including their needs. They described that this process reduced the mental load of creating careful and precise prompts in the initial interactions. As P4 explained, \textit{``The verification process gave me freedom from being so constrained or stressed about making my prompts detailed. I don't always put in a prompt the same way, so it was like guidance, checking to ensure I put in good prompts.''}

\subsubsection{The Need for Contextual Memory for Iterative Rule Verification}

Three participants (P3, P9, P12) suggested improving the rule verification process by enabling the translator to remember past contexts or interactions. Currently, \tool{} only supports single interactions, translating a list of rules based on the initial prompt and resetting the translation process with each new interaction. As a result, if participants wanted to adjust a rule, they had to re-enter their set of instructions with every prompt. They felt that allowing the translator to remember previous rules and iteratively build upon them would be more efficient and helpful in adjusting the rules on the fly and giving feedback to improve individual rules. 

\subsection{Flexibility Sliders Showed Potential to Improve Usefulness of LLMs}
Participants' scores in the \textit{C2 ($\neg$Sliders)} \revision{were significantly lower in usefulness,} and lower in perceived performance, ease of use, and satisfaction compared to \textit{C1 (Full)}. 
While there were no significant differences, the gap in participants' scores for the usefulness measure between \textit{C2 ($\neg$Sliders)} and \textit{C1 (Full)} was larger than in other measures. Our qualitative analysis presents further insights into these results.

\subsubsection{Sliders Were Found Essential for Flexibility in Adaptive Scenarios}
Seven participants (P1, P2, P4, P5, P8, P9, P11) found the sliders particularly useful and necessary in scenarios requiring greater personalization and flexibility in the rules, such as complex scheduling and event planning. They noted that contextual factors, preferences, and priorities often evolve based on user needs, making it essential to consider these variables during planning. In such cases, participants saw the sliders as crucial for managing the complexity of the rules and arriving at a workable solution. However, for tasks such as hospital navigation or recipe planning, which involved stricter rules, participants felt that the sliders were less relevant. Although they appreciated the flexibility sliders provided, they viewed them as secondary to the verification of the translator or model checker. P8 elaborated on this point by \textit{``I think it was just this scenario [hospital navigation], where it felt like these are pretty hard rules. But when you're talking about scheduling, or even personal life, like who gets the kids on what days, there's a lot of flexibility in that, and it would be a lot easier to make decisions. That fits better with life. Like for certain days, you're going to have harder deadlines.''}


\subsubsection{Users Leveraged Sliders to Improve LLM Adaptability and Accuracy}

Participants primarily described two key benefits of the sliders: enabling flexibility in creating plans and helping them understand the capabilities of the system. First, participants highlighted that the sliders allowed for flexibility by adjusting rules based on priorities, preferences, and trade-offs. Ten participants (P1--P6, P8--P11) agreed that the sliders helped them reflect their specific needs, such as safety concerns, reasonable compromises, and personal preferences. One user described this usefulness as \textit{P3: ``There are factors that maybe the AI might not understand---like, for me, family is really important. The previous version without the sliders seemed to prioritize work over family.''}
Some participants (P2, P6, P7, P11) felt that the sliders allowed them to effectively communicate their priorities to the system, conveying their nuances and preferences. One user, P11, described their use of the sliders with the following example, \textit{P11: ``I set rules one and two at 100\%, extremely strict, because I wanted to prioritize patient safety. For rule three, I set it at 70\%---which was about no one traveling without me. I was trying to negotiate, rather than having a binary choice, like in real-life decisions.''}



Additionally, six participants (P1, P3, P4, P6--P8) used the sliders to improve the system's performance by emphasizing the strictness of the rules that the LLM overlooked or was struggling to follow. By increasing the strictness of these rules and relaxing them for well-adhered rules, they aimed to enhance the LLM's output accuracy by directing the system's attention more appropriately. One participant described this intent as \textit{P5: ``The constraints really helped me understand what to emphasize more, based on what the system struggled with. It made me focus on what I wanted the system to prioritize for its own performance when regenerating a plan.''}



\subsubsection{Ambiguity in the Impact of Constraint Strictness}

Four participants (P2, P5, P7, P11) noted that the impact of the slider's strictness was unclear. They were unsure how the specific percentage affected the output or how the system's logic changed based on their specification of strictness. This lack of clarity made it difficult for participants to determine how much to adjust the sliders to reach their planning goal. As a result, some participants felt that it was a trial-and-error process when determining the appropriate level of strictness, forcing them to guess the impact of their choices.
One participant described this ambiguity as \textit{P2: ``Even though I noticed that they have different impacts, and I can try different combinations, it feels a bit up in the air. I don't exactly know what percentage leads to different outcomes, so I wasn't sure how much to change.''}

\subsection{Model Checker Improved Performance, Usefulness, and Satisfaction of LLMs}

Participants' scores in \textit{C1 (Full)} were significantly higher in performance and satisfaction compared to \textit{C3 ($\neg$Translator)} and \textit{C4 (none)}. \revision{In usefulness, \textit{C1 (Full)} was significantly higher than \textit{C2 ($\neg$Slider)}, \textit{C3 ($\neg$Translator)}, and \textit{C4 (none)}.} However, no significant difference was observed in ease of use. Our qualitative findings provide further insights into these results.

\subsubsection{Model Checking for Efficiency and Transparency} 

Nine participants (P2, P3, P5, P6, P8--P12) highlighted that the model checker significantly improved their efficiency by reducing planning time for complex tasks and supporting a constructive trial-and-error process to reach satisfying solutions. They noted that specifying user-centered needs through verified rules and adjusting constraints, followed by the model checker assessing the quality of outputs, greatly improved problem-solving for planning tasks. One participant compared their experience to that of not having model checking, stating, \textit{P10: ``If it hadn't asked to verify things, it would've resulted in more failures, increasing re-do times. The ability to set rules and goals, and then optimize in as few iterations as possible, helped me achieve goals feasibly that would have taken much longer otherwise, you know, figuring out how to bend this way and that.''}

In addition, six participants (P1--P3, P5, P6, P10) emphasized the role of transparency in driving efficiency. The feedback provided by the model checker on system status and errors, along with the input panel for navigating further questions, was particularly useful. One participant described using the input panel for clarification, describing, \textit{P6: ``The input panel was especially useful when I felt like the model checker was assuming something, so then I could ask questions about why it acted that way, and then adjust the rules. So it wasn't that big of a problem.''}

In \textit{C4 (None)} where participants interacted solely with an LLM agent, they reported difficulties in efficiently achieving a correct solution. Seven participants (P2, P4--P6, P10--P12) noted that the LLM often provided ``a'' answer instead of ``the'' answer that best aligned with their needs. They struggled with enforcing rules, as the system did not always capture the specified requirements, leading to inefficiency or failure in achieving the correct outcome. One participant described their experience without the model checker being \textit{P2: ``It felt like it [LLM] was just putting out an answer as fast as possible. I felt more like it was producing an answer, versus trying its best to produce a better answer with its honest, real best effort.''}

Eight participants (P1, P2, P4, P5, P7, P8, P10, P12) also faced challenges with monitoring errors and manually verifying outputs, which added a significant burden, explaining \textit{P10: ``Sometimes speed isn't everything, because all it really did was produce the wrong answer faster. Producing the wrong answer faster just made it more inefficient. I had to keep correcting its logic over and over, which made the process very inefficient.''}
This process raised concerns regarding over-reliance and blindly accepting incorrect results, as another user stated \textit{P4: ``If I wasn't paying as much attention, I would have been possibly just accepted its answer, because I assume that it's an intelligent machine or something like that.''}
Finally, transparency of the system was also an issue as participants struggled to understand the system's logic and rule inputs, leading to skepticism and decreased trust in the system's outputs. One participant described their challenges in navigating the system's decision-making process, \textit{P8: ``When I reviewed the plan, I could quickly see it had missed something, but without knowing exactly what it considered, it was harder to trust. It made me more skeptical about whether it was accounting for all of my priorities.''}

\subsubsection{Feedback from the Model Checker Enabled Creativity in Action Planning}

Eight participants (P1, P3--P5, P8--P10, P12) highlighted that multiple planning attempts, facilitated by rule verification and iterations, allowed them to creatively generate plans that adhered to their predefined rules. Users described how the model checker enabled them to experiment with different levels of constraint strictness and rule adaptations, while ensuring safety through verification. The model checker acted as a safety net, allowing users to be exploratory and creative, which they found helpful in identifying optimal plans. As P10 noted, \textit{``The algorithm now has more options to create plans for me, because it has that determined list of rules that it will be checked against. This allows me, or the LLM, to have more options or creativity for coming up with a plan that is functional for difficult constraints.''}

In addition, participants found that the feedback provided by the model checker on broken rules helped them to gain insight for their next steps, even after unsuccessful attempts. Seven participants (P2--P4, P6, P9, P11, P12) explained how this feedback revealed details that the system missed, highlighted errors, and demonstrated how well the system was interpreting their rules. This understanding enabled them to refine their prompts and adjust constraints to guide the system toward generating the correct solution. Five participants (P1--P3, P5, P12) further emphasized how the feedback revealed alternative solutions they had not previously considered, which helped them compromise, prioritize rules, and understand trade-offs.
P5 illustrated this experience, stating, \textit{``I was able to bounce ideas from its [plan] suggestions and get creative. Like, I never thought you could bring people back to the waiting room and then into the counseling room. That was something I hadn't considered, but it became part of the solution. I learned something from it and added something to my own ideas.''}








\subsubsection{Users Wanted Actionable Suggestions from the Model Checker}
Ten participants (P1--P5, P7, P9--P12) emphasized the importance of the system providing actionable suggestions alongside its planning attempts, which they suggested might enhance usability and efficiency. These suggestions could include adjustments to constraints (\textit{e.g.,} \textit{P12: ``Loosen the time constraint slightly to make the recipe more manageable''}) or guidance on how to rephrase rules to improve model checker comprehension. Users also envisioned the system offering multiple options for resolving issues, allowing them to select the most appropriate adjustment.

One participant suggested that the system highlight broken rules and provide potential fixes. P2 noted, \textit{``What if the system showed you broken rules and said, `Here's how you can fix it,' offering hypotheses for the changes you could try?''} This sentiment was echoed by another participant, who emphasized the importance of prioritization in such suggestions. P3 explained, \textit{``It could inform you about the constraints that are broken and suggest which ones you can adjust, but also warn you about those that are too critical to change, like if someone's safety is at risk.''}

Furthermore, several participants (P3, P5, P6, P9) expressed a desire for the system to take into account the strictness of constraints when offering action plans. P3 elaborated, \textit{``The system could suggest changes to constraints, and I could review the options---maybe rule four isn't that important, so I could go with a solution that adjusts it.''}

\revision{
\subsection{\ours{} Interface Supported Usability}
Our qualitative analysis shows that the design of the front-end interface supported users' interactions with \ours{} in terms of understanding the planning context, applying feedback to user control features, and organizing plans based on user preferences. 

\subsubsection{Understanding Planning Content}
Six participants (P2--P4, P6, P7, P11) highlighted that the \textit{P6: ``mind map layout''} of \ours{} helped them better understand the LLM's reasoning and functionality. \ours{}'s layout organizes key information---such as rules, inputs, outputs, and conflicts---into blocks connected within a visual map. Participants found this compartmentalized structure significantly more intuitive for interpreting planning content compared to conventional text-based LLM interactions.

They explained that text-based interfaces often present a ``wall of text,'' making it difficult to quickly or efficiently extract information about the system's reasoning or conflicts, thereby hindering the system's transparency for user understanding. As one participant noted,
\textit{P3: ``This structure makes it much clearer to see what the plan was, where the conflict happened, and why it occurred. It's all laid out logically, so I can address it right away. With ChatGPT, I'd have to sift through a wall of text and ask multiple follow-up questions just to figure out what went wrong, which takes a lot more effort.''}

\subsubsection{Applying Feedback to User Control Features}
The mind map-based layout was also described to support participants in effectively applying the model checker's feedback. Four participants (P2, P7, P9, P10) described how having all components---the rule translator, flexibility sliders, user's input, planning output, and conflicts from the model checker---in one view and interconnected made it intuitive to apply modifications while monitoring feedback and conflicts in the output. As P9 explained: \textit{``Getting the feedback, I could tweak a slider or update a rule and immediately see how it shifted the output---like adjusting dials on a machine and watching it respond.''}
Three participants (P1, P2, P7) also emphasized that \ours{}'s structure displaying multiple planning iterations in one view helped them track their rule modifications, compare the impact of different rules and adjustments on the LLM's performance, and make their modifications incrementally.

\subsubsection{Organizing Plans Based on User Preferences}
During interactions with \ours{}, multiple participants (P2, P5, P10, P11) used the mind map structure to organize iterations or plans based on their preferences or perceived efficiency. For example, one participant (P5) engaged with the weekly scheduling scenario and described preferring meetings in the morning. They grouped plans with morning meetings into a ``preferred'' category, separating them from plans that scheduled meetings later in the day. They also created a ``less favored'' group for plans where meetings followed their workout sessions, as they disliked feeling sweaty or tired during meetings.
Similarly, another participant (P11), working with the cooking optimization scenario, used the mind map to prioritize plans they found more efficient. For instance, they preferred plans that consolidated ingredient preparation at the beginning rather than doing it separately for each dish, describing as 
\textit{P11: ``That doesn't seem that efficient to me. I like to use my cutting board once and then clean it up. So I prioritized plans like that.''}
Participants described the ability to categorize plans was helpful in selecting or ranking their preferred options and gaining insights into creating the most optimal plan for their needs.
}

\section{Discussion}
\revision{
In this work, we present \ours, which applies formal verification, specifically model checking, to LLMs for complex end-user planning tasks. \ours{} includes three core features---the rule translator, flexibility sliders, and a model checker---and engages users throughout the verification process. Our user study demonstrates that \ours{} enhances users' perceived performance of the LLM, as well as its usefulness, satisfaction, and reliability. Below, we discuss how our findings address the research questions and present design implications for integrating verification processes and user control features into future systems.


\subsection{Formal Verification for Deterministic Boundaries in LLMs}

LLMs have made automated planning more accessible to end-users by removing many of the barriers traditionally associated with planning tools. Existing tools often require users to understand complex formal languages, interpret low-level feedback, build detailed system models, and work within rigid workflows. These challenges are compounded by scalability issues, language barriers, and misalignment with end-user objectives, making them less adaptable to practical, real-world contexts. While LLMs address these accessibility issues for automated planning, their probabilistic nature introduces new risks, including unpredictability in their outputs. This inherent variability can lead to errors and failures, posing notable challenges for ensuring reliability and user confidence \cite{kambhampati2024llms}.

In this work, we aim to combine the strengths of both approaches: LLMs to enhance accessibility of planning tools to end-users, and formal verification methods to ensure safety, reliability, and trustworthiness. Quantitative findings show that LLMs incorporating verification approaches (C1) significantly improves users' perceived performance, usefulness, and satisfaction compared to those without (C4). 
Qualitative insights further illustrate how the model checker, guided by user-defined constraints, effectively aligns the LLM planner's outputs with user needs and goals.
Participants described the model checker as a ``problem solver'' that identified conflicts on their behalf and helped propose solutions, allowing them to achieve their goals more efficiently and reducing concerns about undetected errors. Moreover, participants referred to the model checker as a ``safety net,'' particularly valuable when experimenting with exploratory inputs or modifications related to the rules. By employing an external verification process on plans using user-defined constraints, the model checker alleviates the cognitive burden of manually reviewing constraints and comparing them with the generated outputs, while fostering increased reliability in interactions with the LLM.

These results suggest that formal verification, particularly model checking, can provide deterministic boundaries for the inherently probabilistic nature of LLM systems. By systematically exploring all possible states of a system, model checkers verify whether logical properties are satisfied and, if not, identify violations. This capability allows model checkers to act as external guardrails for LLMs, detecting errors caused by inaccuracies, hallucinations, or misaligned outputs. As the complexity of planning states and constraints increases, such verification becomes essential to ensure the reliable use of LLMs as planning tools.

Looking ahead, the integration of formal verification processes can play an increasingly critical role as LLMs and AI systems are increasingly used for planning contexts \cite{xie2024travelplanner, song2023llm,yao2022react}. As LLMs become increasingly used as tools for beyond planning contexts (\eg personalization \cite{shang2024personalized,ning2024user}), verification methods can enable users to safely and effectively guide, collaborate with, and customize these systems to meet their specific needs. By providing a deterministic mechanism for error detection, formal verification methods can help LLMs adapt to practical, real-world applications while maintaining safety and reliability.

\subsection{User Control with Model Checking for Improved LLM Outputs and User Experience}
\revisionNew{As described above, model checking can be particularly beneficial for LLMs by imposing deterministic boundaries on their probabilistic nature. However, for model checking to effectively support LLMs in achieving personalization, it must acquire user-specific preferences, constraints, and needs. Traditionally, involving users in this specification process has been an arduous task requiring domain-specific expertise. However, with LLMs enabling natural language interactions---such as the translators in \ours{} that convert LTL properties into natural language---users can engage at a higher level without needing prior knowledge of model checking properties or complex programming language concepts. Instead, they can define and refine their specifications in a user-centered, understandable manner.

Thus, combining model checking and LLMs creates a symbiotic relationship: model checking enables formal verification for LLMs, while LLMs lower the technical barrier for users to engage in model checking. This relationship further establishes an environment where users can actively contribute at different points in the decision-making process. Rather than being passive recipients of AI-generated outputs, users can assume an active role in specifying their needs, preferences, and constraints, thereby guiding and refining LLM outputs in an adaptive manner.

The user's role of driving high-level control throughout the system's decision-making process is critical for enhancing both the quality of LLM outputs and the overall user experience. The importance of granting users agency to shape system behavior is well recognized; recent work in LLMs has increasingly focused on enabling human control, such as segmenting queries into sub-tasks for users to specify personal contexts and preferences \cite{ma2024beyond} or adapting outputs based on user feedback \cite{kirk2023past}. However, our insights from \ours{} emphasize that involving users more directly in the system's decision-making process---particularly in the stages of defining system parameters and behavioral factors---can enable more effective and efficient personalization. Compared to involvement solely at the output level, this approach ensures that user needs are clearly defined from the outset, making adaptation more direct and refinement more targeted within a narrowed space.

To fully leverage the benefits of combining LLMs and model checking, system designs should integrate high-level user control at multiple stages of the decision-making process. Future system designs can take inspiration from \ours{}'s approach, which demonstrated effective strategies for high-level user control:

\paragraph{User-defined specifications and iterative refinement} Users defined constraints and preferences for model checking in natural language and refined them iteratively until the rules aligned with their expectations. This early engagement before the system's final output showed effectiveness in reflecting user needs and fostering trust. Rather than limiting personalization to post-hoc feedback on system outputs, which can lead to abstract inference of user preferences, this approach embeds user input in the system's foundation, enabling direct and meaningful refinement.

\paragraph{Flexible endorsement through high-level adjustments} Users interacted with the slider bars to balance constraints based on their priorities. Interestingly, while users did not specify precise numerical values, they intuitively assigned abstract priority levels, which most considered sufficient for preference specification. This suggests that effective user control may not require fine-grained precision but rather a structured way to articulate high-level preferences.

\paragraph{Seamless interaction through intuitive interfaces} The system interface played a crucial role in streamlining user engagement. By abstracting complex model checking and planning processes, it provided a structured yet intuitive workflow, allowing users to easily understand, categorize, and refine their constraints and plans. Ensuring that the planning and verification procedure feels as seamless and accessible as interacting with LLMs is essential to maintaining a positive user experience and lowering the barrier to adoption for end-user planning.

By designing systems that integrate high-level user control throughout different stages of the decision-making process, model checking specifications can be more accurately tailored to user needs, leading to more effective personalization and improved alignment between LLM outputs and user expectations.

}

\subsection{Stages of User Engagement for LLM Verification and Alignment}

Our findings indicate that user engagement with the core features, the rule translator and flexibility sliders, can effectively steer the direction of the LLM and refine it to align with the user's diverse needs and preferences. Based on insights from our findings, we identify two general stages in a user's interaction process with an LLM where user engagement can be beneficial: (1) the initial definition of guidelines and rules for LLM performance, and (2) iterative refinement based on model-checking outputs.

In the initial stage, users can set general guidelines, such as constraints, preferences, or protocols, to align the LLM's performance with their unique needs. Existing research shows that user input on preferences and domain knowledge during initial interactions is important to effectively guide system behavior \cite{kumar2024applications}. Without explicit user input at this stage, LLMs may have difficulty in inferring distinctive user preferences or goals, which are essential for tailoring the system's outputs. This stage can also be particularly effective for user engagement, as users might find it easier to define high-level preferences compared to specifying granular operational details, or they may not always have detailed insights or clarity into their needs and preferences \cite{lee2024ai, simonson2005determinants}. Our results support this notion, as the LLM-generated plans often inspired participants by presenting novel, creative, or efficient approaches they had not previously considered. This highlights a symbiotic relationship where user-defined boundaries enable the system to explore within those parameters, leveraging its computational capabilities to deliver outputs that align or may exceed user expectations.

The second stage of user engagement can occur after the model checker identifies conflicts and provides feedback. By this stage, users have outlined their general preferences and know their interaction goals. What remains unclear are the specific details of their preferences, priorities, or contextual needs, which may vary between users and are challenging for an LLM to infer. However, when presented with a plan or actionable feedback, it can be easier for users to effectively identify and articulate their desired adjustments or more granular preferences. Prior research shows that allowing users to refine and adjust system behavior based on initial feedback improves outcomes and enhances user experience \cite{fang2022framework, schellaert2023your}. Therefore this stage can be particularly valuable for user engagement, as it allows users to refine the LLM's behavior based on personalized intricacies or distinct nuances using tools like the flexibility slider. These refinements can also enable the LLM to capture additional cues and information for future adaptation.

This two-stage approach---initial user input to define the system's scope and subsequent refinement based on feedback---leverages an effective balance between user involvement and the system's autonomy. By allowing the system to independently generate solutions within user-defined boundaries, users can benefit from its ability to propose innovative and comprehensive outputs. Simultaneously, user-driven inputs and refinements can ensure that the system remains responsive to individual needs and evolving goals. Ultimately, this interplay between user input and system capabilities can foster a more effective and user-centered verification process.

\subsection{Design Implications}
Below, we present design implications for incorporating verification processes and user control features into future systems and interaction design.


\subsubsection{Considerations for Integrating Verification Methods into LLMs}

\revisionNew{Our findings demonstrate that formal verification methods, such as model checking, can serve as effective guardrails when applied to LLMs, enhancing perceived performance and user experience. Not limited to model checking, LLM designers should consider integrating various external verification techniques into system designs to ensure reliable verification. Beyond using LLMs for verification, prior work has explored how external verifiers can complement LLM capabilities to address their inherent lack of reliable self-verification. \citet{kambhampati2024llms} introduced a modulo framework, which combines the generative strengths of LLMs with external ``critics'' or verifiers. In this framework, LLMs generate candidate plans and ideas, which are then evaluated by specialized critics leveraging formal domain models and planning algorithms. In other work, \citet{zhang2023controlling} verify LLM outputs by comparing the present state with historical trajectories extracted from a memory module, enabling evaluation and learning. \citet{gou2023critic} employ external tools (\eg knowledge bases, code interpreters, search engines, and calculators) to critique and refine LLM outputs.
Therefore like model checking, LLM designers should explore non-LLM-based verification methods that align with their specific task goals to ensure reliability and effectiveness.}



\subsubsection{Considerations for Incorporating Multiple Dimensions of User Control}

\revisionNew{In \ours{}, both the rule translator and flexibility sliders provide distinct yet complementary dimensions of control: the rule translator allows users to define and refine strict verification boundaries, ensuring alignment with their needs, while the flexibility sliders enable users to adjust the relative weights of rules, adapting the system based on context, evolving preferences, and user priorities. Together, they balance rigid rule definition with nuanced customization to meet diverse user requirements.

Similarly, LLM designers should incorporate appropriate levels and opportunities for user control to optimize system performance, usefulness, and satisfaction. For example, systems could proactively infer user preferences from behavior and engage users for verification, or dynamically adjust the user's autonomy during interactions when the system determines that user control is unnecessary. User control has long been recognized as a critical feature in human-computer interaction, influencing user experience and outcome quality \cite{shneiderman2010designing, nielsen1999designing, harambam2019designing, harper2015putting, andjelkovic2016moodplay}. However, \citet{jin2017different} caution that excessive control can increase cognitive load, emphasizing the need to tailor control to task and user characteristics, such as familiarity and domain knowledge, for balanced usability and effectiveness. Therefore, multi-dimensional control mechanisms should be carefully designed to enable LLM systems to gather richer inputs and seamlessly integrate human knowledge and preferences into their decision-making processes.}

\subsubsection{Considerations for Designing Flexibility Sliders Based on Task and Constraint Characteristics}

Our findings show that the usage pattern for flexibility sliders largely depended on the characteristics of the constraints. When participants worked with organizational or strict rules that allowed little room for negotiation, they used the slider to guide the system to best adhere to all the constraints, ultimately aiming to efficiently obtain an accurate, verified output. In contrast, for constraints reflecting personal preferences or priorities, participants employed the slider as a representation of their values, adjusting it to adaptively align the system's behavior with their evolving needs. 

Designers could adapt flexibility sliders to play various roles, tailored to the task and constraint characteristics. For instance, in workplace contexts that use AI systems for decision-making or plan generation such as the healthcare domain (\eg ensuring diagnosis or treatment plans adhere to medical protocols \cite{hosny2018artificial}) or financial underwriting (\eg creating financial plans while complying with regulatory rules \cite{mayer2020unintended}), sliders can emphasize under-adhered-to rules or prioritize task-specific constraints. Additionally, sliders can allow the worker to explore alternative outcomes, such as loosening a ``no student loans'' rule to assess how it affects a mortgage decision, providing the worker insights for financial advising. Conversely, in personalized contexts like movie recommendations, sliders can enable users to dynamically adjust preferences, such as exploring genres based on their current mood. Thus, future systems should leverage flexibility sliders to support diverse roles, such as facilitating efficient task completion in structured environments or fostering adaptive outputs in more flexible, personalized settings.

\subsubsection{Considerations for Designing Effective Interfaces for LLMs}

Users reported that the interface enhanced their understanding of planning content, facilitated effective feedback and modifications through user control features, and helped organize generated plans to align with their preferences. They particularly appreciated interactive elements such as rule checkboxes and flexibility sliders for providing input beyond text, as well as the mind map-based layout for managing plan generation and incorporating feedback from the model checker. These features improved users' ability to interpret system outputs, reorganize plans efficiently, and gain actionable insights for the next steps.

To design future systems that incorporate verification approaches for LLMs, designers should consider visual, intuitive, and interactive interfaces to enhance usability and satisfaction. Recent studies have emphasized the value of interfaces and visualizations in helping users better understand, organize, and utilize information from LLMs \cite{ma2024beyond, suh2023sensecape, jiang2023graphologue, wang2024farsight}. For complex tasks like planning, where users must manage multiple constraints or variables and compare outputs, text-based interactions alone may be insufficient. Instead, systems could integrate features such as tools for saving and retrieving plans that effectively align with user-defined preferences, drag-and-drop interfaces for reorganizing plan components, timeline views for tracking evolving changes in needs or preferences, or dashboard summaries for visualizing comparisons.

}


\section{Limitations \& Future Work}

While our proposed system offers valuable contributions to integrating formal verification with LLMs, several limitations exist that suggest areas for future improvement. First, the types of temporal constraints available in our template represent a limited subset of potential constraints. Users could benefit from greater flexibility, particularly the ability to define their own temporal constraints, or through enhanced capabilities of LLMs that could be trained or fine-tuned to handle a wider range of temporal constraints.

Additionally, the limitations of the current modeling framework, specifically using PRISM and Stormpy for verification, restrict the types of temporal constraints or logical expressions that can be formulated in LTL. Future research should explore alternative model-checking and formal verification approaches to enhance the expressiveness and applicability of formal verification in conjunction with LLMs.
Future work could also focus on improving the output of the model checker within \ours{} by introducing proactive suggestions and actions. This improvement could be achieved through program repair techniques, such as automatic or interactive repair.

\revision{The evaluation of \ours{} is limited to three scenarios, focusing on planning tasks that end-users can describe in natural language. Future work could explore and test the effectiveness of \ours{} in various domain-specific planning tasks, such as those in healthcare or manufacturing.

Finally, our sample size is limited to 12 participants. Larger scale studies can be conducted to validate and expand the results and reveal additional insights beyond those identified in this paper.}




\section{Conclusion}
This paper introduces \ours, a system that integrates formal verification techniques with LLMs to enhance their reliability and usability for end-user planning tasks. Our evaluation shows that the core features of \ours---the rule translator, flexibility sliders, and model checker---improved users' perceptions of performance, usability, satisfaction, and reliability in LLM outputs. These findings emphasize the value of incorporating formal verification methods in LLMs for everyday users, providing rigidity and deterministic boundaries to mitigate the probabilistic nature of LLMs, making them more reliable for planning tasks. The integration of user-controlled flexibility in verification further enhanced creativity in plan generation and aligned outputs with personal preferences and evolving contexts. Finally, our results underscore the importance of user control in the model-checking process, which significantly improves the reliability and usability of LLM outputs. Based on these insights, \ours{} offers valuable implications for LLMs as end-user planning tools,  highlighting the need for verification methods and user control features to ensure reliability, user-centered adaptability, and alignment with complex real-world needs.

\begin{acks}
We thank the reviewers for their helpful comments. This work was supported by the National Science Foundation awards 1925043 and 2152163. This research was also partially supported by the U.S. Naval Research Laboratory (NRL) and an NRC Postdoctoral Research Associateship awarded to DP at NRL. The views and conclusions contained herein are those of the authors and should not be interpreted as necessarily representing the official policies, either expressed or implied, of the U.S. Navy.
\end{acks}
\balance
\bibliographystyle{ACM-Reference-Format}
\bibliography{bibliography}

\end{document}